\documentclass[twocolumn,aps,prl,showpacs]{revtex4}
\begin{document}
\title{Violating Bell's Inequality Beyond Cirel'son's Bound}
\author{Ad\'{a}n Cabello}
\email{adan@us.es}
\affiliation{Departamento de F\'{\i}sica Aplicada II,
Universidad de Sevilla, 41012 Sevilla, Spain}
\date{\today}


\begin{abstract}
Cirel'son inequality states that the absolute value of the
combination of quantum correlations appearing in the
Clauser-Horne-Shimony-Holt (CHSH) inequality is bound by $2 \sqrt
2$. It is shown that the correlations of two qubits belonging to a
three-qubit system can violate the CHSH inequality beyond $2 \sqrt
2$. Such a violation is not in conflict with Cirel'son's
inequality because it is based on postselected systems. The
maximum allowed violation of the CHSH inequality, 4, can be
achieved using a Greenberger-Horne-Zeilinger state.
\end{abstract}


\maketitle


Bell's theorem \cite{Bell64} has been described as ``the most
profound discovery of science'' \cite{Stapp75}. It states that,
according to quantum mechanics, the value of a certain combination
of correlations for experiments on two distant systems can be
higher than the highest value allowed by any local-realistic
theory of the type proposed by Einstein, Podolsky, and Rosen
\cite{EPR35}, in which local properties of a system determine the
result of any experiment on that system. The most commonly
discussed Bell inequality, the Clauser-Horne-Shimony-Holt (CHSH)
inequality \cite{CHSH69}, states that in any local-realistic
theory the absolute value of a combination of four correlations is
bound by 2. Cirel'son's inequality \cite{Cirelson80} shows that
the combination of {\em quantum} correlations appearing in the
CHSH inequality is bound by $2 \sqrt{2}$ (Cirel'son's bound). It
is widely believed that ``[q]uantum theory does not allow any
stronger violation of the CHSH inequality than the one already
achieved in Aspect's experiment \cite{ADR82} [$2 \sqrt{2}$]''
\cite{Peres93}. However, it has been shown that exceeding
Cirel'son's bound is not forbidden by relativistic causality
\cite{PR94}. Therefore, an intriguing question is why is the CHSH inequality
not violated {\em more}. Here it is shown that,
for three-qubit systems (that is, systems composed by three
two-level quantum particles), the correlation functions of {\em
two} suitably postselected qubits violate the CHSH inequality
beyond Cirel'son's bound and that this violation can even reach 4,
the maximum value allowed by the definition of correlation.


To introduce the CHSH inequality, let us consider systems with two
distant particles $i$ and $j$. Let $A$ and $a$ ($B$ and $b$) be
physical observables taking values $-1$ or $1$ referring to local
experiments on particle $i$ ($j$). The correlation $C(A,B)$ of $A$
and $B$ is defined as
\begin{eqnarray}
C(A,B) & = & P_{AB}(1,1)-P_{AB}(1,-1) \nonumber \\
 & & -P_{AB}(-1,1)+P_{AB}(-1,-1),
\end{eqnarray}
where $P_{AB}(1,-1)$ denotes the joint probability of
obtaining $A=1$ and $B=-1$ when $A$ and $B$ are measured. In any
local-realistic theory, that is, in any theory in which local
variables of particle $i$ ($j$) determine the results of local
experiments on particle $i$ ($j$), the absolute value of a
particular combination of correlations is bound by $2$:
\begin{eqnarray}
& & |C(A,B) - m\,C(A,b) \nonumber \\
& & -n\, C(a,B) - m n\,C(a,b)| \le 2,
\label{CHSH}
\end{eqnarray}
where $m$ and $n$ can be either $-1$ or 1. The CHSH inequality
(\ref{CHSH}) holds for any local-realistic theory, whatever the
values of $m$ and $n$ are, in the allowed set, $\{-1,1\}$.

The bound 2 in inequality (\ref{CHSH}) can be easily derived as follows:
In a local-realistic theory, for any individual system, the observables
$A$, $a$, $B$, and
$b$ have predefined values $v_A$, $v_a$, $v_B$, and $v_b$, either $-1$ or 1.
Therefore, for an individual system the combination of correlations
appearing in (\ref{CHSH}) can be calculated as
\begin{equation}
v_B (v_A-n\,v_a)-m\,v_b (v_A+n\,v_a),
\end{equation}
which is either $-2$ or 2, because one of the expressions
between parentheses is
necessarily zero and the other is either $-2$ or $2$.
Therefore, the absolute value of
the corresponding averages is bound by $2$, q.e.d.

For a two-particle system in a quantum pure state described by a
vector $\left| \psi \right\rangle$, the quantum correlation of $A$
and $B$ is defined as
\begin{equation}
C_Q(A,B)=\left\langle \psi \right| \hat A
\hat B \left| \psi \right\rangle,
\end{equation}
where $\hat A$ and $\hat B$ are
the self-adjoint operators which represent observables $A$ and
$B$. For certain choices of $\hat A$, $\hat a$, $\hat B$, $\hat
b$, and $\left| \psi \right\rangle$, quantum correlations violate
the CHSH inequality \cite{CHSH69}. Therefore, no local-realistic
theory can reproduce the predictions of quantum mechanics
\cite{Bell64}.


Later on, Cirel'son \cite{Cirelson80} demonstrated that for a
two-particle system the absolute value of the combination of {\em
quantum} correlations equivalent to those appearing in the CHSH
inequality (\ref{CHSH}) is bound by $2 \sqrt{2}$,
\begin{eqnarray}
& & |C_Q(A,B) - m\,C_Q(A,b) \nonumber \\
& & -n\, C_Q(a,B) - m n\,C_Q(a,b)| \le 2 \sqrt{2}.
\label{Cirelson}
\end{eqnarray}
Cirel'son's bound can be easily derived as follows \cite{Landau87}:
Consider the operator with the same structure as the
combination which appears in inequality (\ref{Cirelson}),
\begin{equation}
{\hat C}={\hat A} {\hat B}-m\,{\hat A}\,{\hat b}-n\,{\hat a} {\hat B}
-mn\,{\hat a}\,{\hat b}.
\end{equation}
If ${\hat A}^2={\hat a}^2={\hat B}^2={\hat b}^2=
1\!\!{\mbox I}$, where $1\!\!{\mbox I}$ is the identity operator,
\begin{equation}
{\hat C}^2 = 4\,1\!\!{\mbox I}
-mn\,[{\hat A},{\hat a}]\,[{\hat B},{\hat b}].
\end{equation}
Since for all ${\hat F}$ and ${\hat G}$ bounded operators,
\begin{equation}
||\,[{\hat F},{\hat G}]\,|| \le ||{\hat F}{\hat G}||+||{\hat G}{\hat F}||
\le 2\,||{\hat F}||\,||{\hat G}||,
\end{equation}
then $|| {\hat C}^2 || \le 8$, or $|| {\hat C} || \le 2 \sqrt{2}$, q.e.d.

Different derivations of this bound can be found in \cite{BMR92,CB96}.
Violations of the CHSH
inequality (\ref{CHSH}) by $2 \sqrt{2}$ can be obtained with pure
\cite{CHSH69} or mixed states \cite{BMR92}.


Popescu and Rohrlich \cite{PR94} raised the question whether
relativistic causality could restrict the violation of the CHSH
inequality to $2 \sqrt{2}$ instead of $4$, which would be the
maximum bound allowed if the four correlations in the CHSH
inequality (\ref{CHSH}) were independent. They prove this
conjecture false \cite{PR94} by defining a contrived correlation
function which satisfies relativistic causality while still
violating the CHSH inequality by the maximum value 4.


Here I shall show that violations of the CHSH inequality beyond
the $2 \sqrt{2}$ bound can be naturally obtained using only the
predictions of quantum mechanics. This does not entail a violation
of Cirel'son's inequality but a violation of the CHSH inequality
beyond Cirel'son's bound. To understand the difference, let us
consider three identical brothers. Every morning each of them
takes a bus in London. One goes to Aylesbury, other to Brighton,
and the third to Cambridge. Two of them wear white coats and the
other wears a black one. We are interested in the correlations
between the experiments on two of them. Then, the first step is to
define which two. One possibility is to choose those brothers
arriving in Aylesbury and Brighton. Other possibility is to choose
those wearing white coats, regardless of their destination. Both
possibilities are legitimate in a theory in which both procedures
used for selecting pairs are related to predefined properties.
According to Einstein, Podolsky, and Rosen \cite{EPR35}, a local
system is assumed to have a predefined property if we can predict
with certainty the value of that property from the results of
experiments on distant systems. Therefore, for a local-realistic
theory, both procedures described above for selecting pairs would
be legitimate. If we are interested in the correlations between
the experiments on the two brothers with white coats, we can see
whether the coat of the brother arriving in Cambridge is black. If
this is the case, we can legitimate conclude that the other two
brothers wear white coats. However, in quantum mechanics it is not
meaningful to assume that some physical observables have
predefined values before the measurements are made. Therefore,
such an inference is not permitted.

The key for the understanding of our approach is to realize that, in searching
for violations of the CHSH inequality
(which is derived assuming local-realism,
without any mention of quantum mechanics),
one is not limited to
studying only the correlations of systems prepared in a quantum state, as
in Cirel'son's inequality, but rather that
one can use the correlations predicted
by quantum mechanics for different subsets of systems
previously prepared in a quantum state.
Therefore, one can use a procedure like the one
described above for
selecting pairs. However, one cannot do this if we are interested
in violations of Cirelson's inequality, which is valid for
systems prepared in a quantum state (without any further postselection).

The important point for physics
is not whether a quantum state violates the
CHSH inequality but rather whether the predictions
of quantum mechanics violate
the CHSH inequality and the extent of this violation.
We will show
that the correlations of a postselected subsystem of
a three-qubit system prepared in a
Greenberger-Horne-Zeilinger (GHZ) state
\cite{GHZ89,Mermin90ab,GHSZ90} as described by quantum mechanics
allow the maximum violation.

Let us consider systems of three distant
qubits prepared in the GHZ state:
\begin{equation}
\left| \Psi \right\rangle = {1 \over \sqrt{2}} \left( {\left| +++
\right\rangle + \left| --- \right\rangle } \right),
\label{GHZ}
\end{equation}
where $+$ and $-$ denote, respectively, spin-up and spin-down in
the $y$ direction. For each three-qubit system prepared in the state
(\ref{GHZ}), let us denote as qubits $i$ and $j$ those giving the
result $-1$ when measuring the spin in the $z$ direction on all
three qubits; the third qubit will be denoted as $k$. If all three
qubits give the result $1$, qubits $i$ and $j$ could be any pair
of them. Since no other combination of results is allowed for
state (\ref{GHZ}), qubits $i$ and $j$ are well defined for every
three-qubit system.
Generally, qubits $i$
and $j$ will be in a different location for each three-qubit
system. For instance, if we denote the three possible locations as
1, 2, and 3, in the first three-qubit system, qubits $i$ and $j$
could be in locations 1 and 2; in the second three-qubit system,
they could be in locations 1 and 3, etc.
However, we can force qubits $i$ and $j$ to be those
in $1$ and $2$, just by measuring the spin in the $z$ direction on
the qubit in $3$ and then selecting only those events in which
the result of this measurement is $1$.

We are interested in the
correlations between two observables $A$ and $a$ of qubit $i$ and
two observables $B$ and $b$ of qubit $j$. In particular, let us
choose $A=Z_i$, $a=X_i$, $B=Z_j$, and $b=X_j$, where $Z_q$ and
$X_q$ are the spin of qubit $q$ along the $z$ and $x$ directions,
respectively. The particular CHSH inequality (\ref{CHSH}) we are
interested in is the one in which $m=n=x_k$, where $x_k$ is one of
the possible results, $-1$ or $1$ (although we do not know which one),
of measuring $X_k$. With this choice we obtain the CHSH
inequality:
\begin{eqnarray}
& & |C(Z_i,Z_j) - x_k \,C(Z_i,X_j) \nonumber \\
& & -x_k\, C(X_i,Z_j) - C(X_i,X_j)| \le 2,
\label{CHSH23}
\end{eqnarray}
which holds for any local-realistic theory, regardless of the
particular value, either $-1$ or $1$, of $x_k$. Now let us use
quantum mechanics to calculate the four correlations appearing in
(\ref{CHSH23}). By the definition of qubits $i$ and $j$, and
taking into account that state (\ref{GHZ}) is an eigenstate of the
self-adjoint operator $\hat Z_i \hat Z_j \hat Z_k$ with eigenvalue
$1$, we obtain
\begin{equation}
C(Z_i,Z_j)= 1, \end{equation} since the only possible results are
$Z_i=Z_j=1$ and $Z_i=Z_j=-1$. By taking into account that state
(\ref{GHZ}) is an eigenstate of $\hat Z_i \hat X_j \hat X_k$ with
eigenvalue $-1$, we obtain
\begin{equation}
C(Z_i,X_j)= -x_k,
\label{czx}
\end{equation}
since the only possible results are $Z_i=1$, $X_j=-x_k$ and
$Z_i=-1$, $X_j=x_k$. By taking into account that state (\ref{GHZ})
is an eigenstate of $\hat X_i \hat Z_j \hat X_k$ with eigenvalue
$-1$, we obtain
\begin{equation}
C(X_i,Z_j)= -x_k,
\label{cxz}
\end{equation}
since the only possible results are $X_i=x_k$, $Z_j=-1$ and
$X_i=-x_k$, $Z_j=1$. Finally, by the definition of qubit $k$ as
the one in which $z_k=1$, and taking into account that state
(\ref{GHZ}) is an eigenstate of $\hat X_i \hat X_j \hat Z_k$ with
eigenvalue $-1$, we obtain
\begin{equation}
C(X_i,X_j)= -1,
\end{equation}
since the only possible results are $X_i=-X_j=1$ and
$X_i=-X_j=-1$. Therefore, the left-hand side of inequality
(\ref{CHSH23}) is 4, which is the {\em maximum} value allowed by
the definition of correlation. Other choices of three-qubit
entangled quantum states and observables lead to violations of the
CHSH inequality in the $2 \sqrt{2}$ to $4$ range.


This result opens the possibility of
using sources of quantum entangled states of three or more
particles \cite{BPDWZ99} to experimentally test \cite{PBDWZ00}
local realism using not only proofs of Bell's theorem without
inequalities \cite{GHZ89,Mermin90ab,GHSZ90} or Bell inequalities
involving correlations between three or more particles
\cite{Mermin90c,RS91}, but also the CHSH
inequality (\ref{CHSH23}).

However, it must be stressed that, since the
correlations (\ref{czx}) and (\ref{cxz}) between qubits $i$ and $j$
depend on qubit $k$, the
experimental test cannot be simply
a test on, for instance, those pairs arriving in locations 1 and 2
when a particular measurement on the qubit arriving in 3 gives a
particular result, but,
as we shall see below,
it requires treating all three qubits in a completely
symmetrical way.

On the other hand,
in real experiments using three qubits, the experimental data
consist on the number of coincidences (that is, of simultaneous
detections by three detectors) $N_{ABC} \left( {a,b,c} \right)$
for various observables $A$, $B$, and $C$. This number is
proportional to the corresponding joint probability, $P_{ABC}
\left( {a,b,c} \right)$.

Therefore, in order to make inequality
(\ref{CHSH23}) useful for real experiments,
we must firstly translate it
into the language of joint probabilities and then
we must show how the joint probabilities
of qubits $i$ and $j$ are
related to the probabilities of coincidences of
qubits arriving in 1, 2, and 3.

For the first step, it is useful to
note that, by assuming physical locality (that is, that the
expected value of any local observable cannot be affected by
anything done to a distant particle), the CHSH inequality
(\ref{CHSH23}) can be transformed into a more convenient
experimental inequality \cite{CH74,Mermin95}:
\begin{eqnarray}
-1\!\!& \le &\!\!P_{Z_i Z_j} \left( {-1,-1} \right) - P_{Z_i X_j} \left(
{-1,-x_k} \right) \nonumber \\
& &\!\!-P_{X_i Z_j} \left( {-x_k,-1}
\right)-
 P_{X_i X_j} \left( {x_k,x_k} \right) \le 0.
\label{CH23}
\end{eqnarray}
The bounds $l$ of inequalities (\ref{CHSH}) and (\ref{CHSH23}) are
transformed into the bounds $(l-2)/4$ of inequality (\ref{CH23}).
Therefore, the local-realistic bound in (\ref{CH23}) is 0,
Cirel'son's bound is $(\sqrt{2}-1)/2$, and the maximum value is
$1/2$. For qubits $i$ and $j$ of a system in the state (\ref{GHZ}),
\begin{equation}
P_{Z_i Z_j} \left( {-1,-1} \right)=3/4,
\label{pzz}
\end{equation}
since, in the state (\ref{GHZ}), the four possible results satisfying
$z_i z_j z_k=1$ (where $z_i$ denotes the result of measuring
$Z_i$, etc.) have probability $1/4$ and in three of them $-1$
appears twice;
\begin{equation}
P_{Z_i X_j} \left( {-1,-x_k} \right)=0,
\label{pzx}
\end{equation}
since, in the state (\ref{GHZ}), $z_i x_j x_k=-1$;
\begin{equation}
P_{X_i Z_j} \left({-x_k,-1} \right)=0,
\label{pxz}
\end{equation}
since, in the state (\ref{GHZ}), $x_i z_j x_k=-1$;
\begin{equation}
P_{X_i X_j} \left( {x_k,x_k} \right)=1/4,
\label{pxx}
\end{equation}
since, in the state (\ref{GHZ}), both results $x_i=x_j=x_k=1$ and
$x_i=x_j=x_k=-1$ have probability $1/8$. Therefore, as expected,
the maximum allowed violation of inequality (\ref{CH23}) occurs
for the same choices in which the maximum violation of the CHSH
inequality (\ref{CHSH23}) does.

The second step consists on
showing how the
four joint probabilities (\ref{pzz})--(\ref{pxx}) are
related to the probabilities of coincidences in an experiment with
three spatial locations, 1, 2, and 3.
As can easily be seen,
\begin{eqnarray}
P_{Z_i Z_j} \left( {-1,-1} \right) & = & P_{Z_1 Z_2 Z_3} \left(
{1,-1,-1} \right) \nonumber \\
& & +P_{Z_1 Z_2 Z_3} \left({-1,1,-1} \right) \nonumber \\
& & +P_{Z_1 Z_2 Z_3} \left({-1,-1,1} \right) \nonumber \\
& & +P_{Z_1 Z_2 Z_3} \left({-1,-1,-1} \right),
\label{PZZ}
\end{eqnarray}
where, in the state (\ref{GHZ}), the first three probabilities in the
right-hand side of (\ref{PZZ}) are expected to be $1/4$ and the
fourth is expected to be zero. On the other hand,
$P_{Z_i X_j} \left( {-1,-x_k}
\right)$ and $P_{X_i Z_j} \left( {-x_k,-1} \right)$ are both less
than or equal to
\begin{eqnarray}
& &\!\!\!\!P_{Z_1 X_2 X_3} \left( {-1,1,-1} \right)+P_{Z_1 X_2 X_3}
\left( {-1,-1,1} \right) \nonumber \\
& &\!\!\!\!+P_{X_1 Z_2 X_3} \left(
{1,-1,-1} \right)+P_{X_1 Z_2 X_3} \left( {-1,-1,1} \right)
\nonumber \\
& &\!\!\!\!+P_{X_1 X_2 Z_3} \left( {1,-1,-1} \right)+P_{X_1 X_2 Z_3} \left(
{-1,1,-1} \right),
\label{sum}
\end{eqnarray}
where, in the state (\ref{GHZ}), the six probabilities in (\ref{sum})
are expected to be zero. Finally,
\begin{eqnarray}
P_{X_i X_j} \left( {x_k,x_k} \right) & = & P_{X_1 X_2 X_3} \left(
{1,1,1} \right) \nonumber \\
& & +P_{X_1 X_2 X_3} \left(
{-1,-1,-1} \right),
\label{PXX}
\end{eqnarray}
where, in the state (\ref{GHZ}), the two probabilities in the
right-hand side of (\ref{PXX}) are expected to be $1/8$.

The experimental data of previous tests using three-photon systems
prepared in a GHZ state \cite{PBDWZ00} or
possible new experiments over large distances with spacelike
separated randomly switched measurements \cite{WJSHZ98} or with
three-ion systems and almost-perfect detectors \cite{RKVSIMW01}
could experimentally confirm this violation of
local realism predicted by quantum mechanics.\\


I thank J. L. Cereceda and K. Svozil for comments and the Junta de
Andaluc\'{\i}a grant FQM-239 and the Spanish Ministerio de Ciencia
y Tecnolog\'{\i}a grants BFM2000-0529 and BFM2001-3943 for
support.



\begin{thebibliography}{99}

\bibitem{Bell64} J.S. Bell,
Physics (Long Island City, NY) {\bf 1}, 195 (1964).

\bibitem{Stapp75} H.P. Stapp,
Nuovo Cimento Soc. Ital. Fis. {\bf 29B}, 270 (1975).

\bibitem{EPR35} A. Einstein, B. Podolsky, and N. Rosen,
Phys. Rev. {\bf 47}, 777 (1935).

\bibitem{CHSH69} J.F. Clauser, M.A. Horne, A. Shimony, and R.A. Holt,
Phys. Rev. Lett. {\bf 23}, 880 (1969).

\bibitem{Cirelson80} B.S. Cirel'son,
Lett. Math. Phys. {\bf 4}, 93 (1980).

\bibitem{ADR82} A. Aspect, J. Dalibard, and G. Roger,
Phys. Rev. Lett. {\bf 49}, 1804 (1982).

\bibitem{Peres93} A. Peres,
{\em Quantum Theory: Concepts and Methods} (Kluwer, Dordrecht,
1993), p. 174.

\bibitem{PR94} S. Popescu and D. Rohrlich,
Found. Phys. {\bf 24}, 379 (1994).

\bibitem{Landau87} L.J. Landau,
Phys. Lett. A {\bf 120}, 54 (1987).

\bibitem{BMR92} S.L. Braunstein, A. Mann, and M. Revzen,
Phys. Rev. Lett. {\bf 68}, 3259 (1992).

\bibitem{CB96} A. Chefles and S.M. Barnett,
J. Phys. A {\bf 29}, L237 (1996).

\bibitem{GHZ89} D.M. Greenberger, M.A. Horne, and A. Zeilinger,
in {\em Bell's Theorem, Quantum Theory, and Conceptions of the
Universe}, edited by M. Kafatos (Kluwer, Dordrecht, 1989), p. 69.

\bibitem{Mermin90ab} N.D. Mermin,
Phys. Today {\bf 43}, No. 6, 9 (1990);
Am. J. Phys. {\bf 58}, 731 (1990).

\bibitem{GHSZ90} D.M. Greenberger, M.A. Horne, A. Shimony, and
A. Zeilinger,
Am. J. Phys. {\bf 58}, 1131 (1990).

\bibitem{BPDWZ99} D. Bouwmeester, J.-W. Pan, M. Daniell, H. Weinfurter,
and A. Zeilinger,
Phys. Rev. Lett. {\bf 82}, 1345 (1999).

\bibitem{PBDWZ00} J.-W. Pan, D. Bouwmeester, M. Daniell, H. Weinfurter,
and A. Zeilinger,
Nature (London) {\bf 403}, 515 (2000).

\bibitem{Mermin90c} N.D. Mermin,
Phys. Rev. Lett. {\bf 65}, 1838 (1990).

\bibitem{RS91} S.M. Roy and V. Singh,
Phys. Rev. Lett. {\bf 67}, 2761 (1991).

\bibitem{CH74} J.F. Clauser and M.A. Horne,
Phys. Rev. D {\bf 10}, 526 (1974).

\bibitem{Mermin95} N.D. Mermin,
Ann. N. Y. Acad. Sci. {\bf 755}, 616 (1995).

\bibitem{WJSHZ98} G. Weihs, T. Jennewein, C. Simon, H. Weinfurter,
and A. Zeilinger,
Phys. Rev. Lett. {\bf 81}, 5039 (1998).

\bibitem{RKVSIMW01} M.A. Rowe, D. Kielpinski, V. Meyer,
C.A. Sackett, W.M. Itano, C. Monroe, and D.J. Wineland,
Nature (London) {\bf 409}, 791 (2001).

\end{thebibliography}
\end{document}